\documentclass[preprint,aps]{revtex4}
\begin{document}
\title{Levy-Lieb constrained-search formulation as a minimization of the correlation functional.}
\author{Luigi Delle Site}
 \email{dellsite@mpip-mainz.mpg.de}
\affiliation{Max-Planck-Institute for Polymer Research\\
Ackermannweg 10, D 55021 Mainz Germany.}

\begin{abstract}
The constrained-search formulation of Levy and Lieb, which formally defines the exact Hohenberg-Kohn functional for any $N$-representable electron density, is here shown to be equivalent to the minimization of the correlation functional with respect to the $N-1$ conditional probability density, where $N$ is number of electrons of the system. The consequences and implications of such a result are here analyzed and discussed via a practical example. 
\end{abstract}
\maketitle
\section{Introduction}
The Hohenberg-Kohn (HK) theorem \cite{hkt} has opened new perspectives to the calculations of electronic-based properties of condensed matter \cite{parryang}, and, an aspect often disregarded, has given profound new insights into the general understanding of quantum mechanics. In fact the $3N$-dimensional Schr\"{o}dinger problem for the ground state of an electronic system:
\begin{equation}
H_{N}\psi({\bf r}_{1},...{\bf r}_{2})=E_{0}\psi({\bf r}_{1},...{\bf r}_{2}); H_{N}=\sum_{i=1,N}(-\frac{1}{2}\nabla_{i}^{2})+\sum_{i=1,N}v({\bf r}_{i})+\sum_{i<j}\frac{1}{r_{ij}}
\label{eqsch}
\end{equation} 
where $v({\bf r}_{i})$ is the external potential, $\sum_{i<j}\frac{1}{r_{ij}}$ is the electron-electron Coulomb term, $E_{0}$ is the energy of the ground state and $\psi({\bf r}_{1},...{\bf r}_{N})$ is the $3N$-dimensional antisymmetric ground state wavefunction \cite{au}, is transformed into a ''manageable'' variational problem in three dimensions where the central role is played by the electron density: $\rho({\bf r})=N\large\int_{\Omega_{N-1}}\psi^{*}({\bf r},{\bf r}_{2},.....{\bf r}_{N})\psi({\bf r},{\bf r}_{2},.....{\bf r}_{N})d{\bf r}_{2}....d{\bf r}_{N}$, where $\Omega_{N-1}$ is the $N-1$ spatial domain. In explicit terms the variational problems is written as:
\begin{equation}
E_{0}=Min_{\rho}E[\rho]
\label{eqvar}
\end{equation}
where $\large\int_{\Omega}\rho({\bf r})d{\bf r}=N$ ($\Omega$ being the spatial domain of definition) and $E[\rho]=T[\rho]+V_{ee}[\rho]+V_{ext}[\rho]$ is the energy functional composed respectively by the kinetic, electron-electron potential and the external potential functional.
However in its original formulation the HK theorem and the related variational problem have got a restricted field of applicability; it is valid only if the electron density $\rho({\bf r})$ is $v$-representable, that is if $\rho({\bf r})$ is the density corresponding to an antisymmetric wavefunction of the ground-state of an Hamiltonian of the form of Eq.\ref{eqsch}. It follows that the correct formulation of the variational problem becomes:
\begin{equation}
E_{0}=Min_{\rho}E_{v}[\rho]
\label{eqvr}
\end{equation}
where $v$ refers to the $v$-representability of $\rho({\bf r})$.
As discussed in Ref.\cite{parryang}, there are no general conditions for a density to be $v$-representable and this makes the use of the HK theorem and its associated variational principle not practical. A generalization of the HK theorem which does not require $\rho({\bf r})$ to be $v$-representable was found, in parallel, by M.Levy \cite{levy1} and E.Lieb \cite{lieb1} and it is usually known as the {\bf Levy constrained-search formulation} or {\bf Levy-Lieb constrained-search formulation} \cite{ll}; in this paper we adopt the latter terminology. We also notice that recently P.Ayers \cite{ayers2} has further clarified this concept and developed an axiomatic treatment of the Hohenberg-Kohn functional.
In the following we briefly describe the crucial aspects of the abovementioned approach which are relevant for the current work.
The starting point of the theory is the distinction between the ground state wavefunction, $\psi$, and a wavefunction $\psi_{\lambda}$ that also integrates to the ground state electron density $\rho({\bf r})$.
Since $\psi$ is the ground state wavefunction, we have:
\begin{equation}
\left<\psi_{\lambda}\left|H_{N}\right|\psi_{\lambda}\right>\ge\left<\psi\left|H_{N}\right|\psi\right>=E_{0}.
\label{newvar}
\end{equation}
Taking into account that $V_{ext}[\rho]$ is a functional of $\rho$ only, Eq.\ref{newvar} can be written as:
\begin{equation}
 \left<\psi_{\lambda}\left|T+V_{ee}\right|\psi_{\lambda}\right>\ge\left<\psi\left|T+V_{ee}\right|\psi\right>
\label{snewvar}
\end{equation}
where $T$ and $V_{ee}$ are respectively the kinetic and Coulomb electron-electron operator as defined in Eq.\ref{eqsch}.
The meaning of Eq.\ref{snewvar} is that $\psi$ is the wavefunction that minimize the kinetic plus the electron-electron repulsion energy and integrates to $\rho$. It follows that the initial variational problem of Eq.\ref{eqvar} can be transformed in a double hierarchical minimization procedure which formally allows for searching among all the $\rho$'s which are $N$-representable, i.e. it can be obtained from some antisymmetric wavefunction; this is a condition which is much weaker and more controllable than the $v$-representability. In explicit terms such a formulation is written as:
\begin{equation}
E_{0}=Min_{\rho}\left[Min_{\psi_{\lambda}\to \rho}\left<\psi_{\lambda}\left|T+V_{ee}\right|\psi_{\lambda}\right>+\large\int v({\bf r})\rho({\bf r})d{\bf r}\right].
\label{eq1}
\end{equation}
The inner minimization is restricted to all wave functions $\psi_{\lambda}$ leading to $\rho({\bf r})$, while the outer minimization searches over all the $\rho$'s which integrate to $N$. The original HK formulation can then be seen as a part of this new one once its universal functional, $F[\rho]=\left<\psi\left|T+V_{ee}\right|\psi\right>$ is written as:
\begin{equation}
F[\rho]=Min_{\psi\to \rho}\left<\psi\left|T+V_{ee}\right|\psi\right>.
\label{eq2}
\end{equation}
The purpose of this work is to show that $F[\rho]$ can be determined solely by a minimization with respect to the $N-1$  conditional probability density of the electron correlation functional. This latter will be shown to be composed by the non local Fisher information functional \cite{fisher} and the electron-electron two-particle Coulomb term.
The advantage of this representation is manifold; it further clarifies the connection of electronic properties to the Fisher theory and shows that the knowledge of such a functional is the crucial ingredient in density functional based approaches; it also identifies the Weizsacker kinetic term, $\int\frac{|\nabla\rho({\bf r})|^{2}}{\rho({\bf r})}d{\bf r}$, as necessary component of the universal functional $F[\rho]$ and, in practical terms, offers an objective criterion of evaluation of ''approximate'' exchange and correlation functional, i.e. among two functionals, the physically better founded is the ''smaller'' one. In order to show the practical aspects of our idea we illustrate a potential application.
\section{The new representation}
Before writing the functional in the conditional probability density formalism, we need to define such a quantity.
Let us consider a generic fermionic wavefunction $\psi({\bf r}_{1},....{\bf r}_{N})$, for simplicity we consider a real wavefunction, but the extension to a complex one can be also done \cite{lui1}; we do not consider the spin dependence explicitly,
however this will not influence the main conclusions. Then the $N$-particle probability density is \cite{sears,ayers1}:
\begin{equation}
N\psi^{*}({\bf r}_{1},....{\bf r}_{N})\psi({\bf r}_{1},....{\bf r}_{N})=\Theta^{2}({\bf r}_{1},....,{\bf r}_{N})
\end{equation}
and this can formally decomposed as \cite{sears,ayers1}:
\begin{equation}
\Theta^{2}({\bf r}_{1},....,{\bf r}_{N})=\rho({\bf r}_{1})f({\bf
  r}_{2},.......,{\bf r}_{N}/{\bf r}_{1})
\label{eq4}
\end{equation}
where $\rho({\bf r}_{1})$ is the one particle probability density (normalized to $N$)
 and $f({\bf
  r}_{2},.......,{\bf r}_{N}/{\bf r}_{1})$ is the
$N-1$ electron conditional (w.r.t. ${\bf r}_{1}$) probability density, i.e. the
probability density of finding an $N-1$ electron configuration, $C({\bf
  r}_{2},.......,{\bf r}_{N})$, for a given fixed
value of ${\bf r}_{1}$. The function $f$ satisfies the following properties:
\begin{eqnarray}
{\it (i)}~~~~~~\large\int_{\Omega_{N-1}}f({\bf
  r}_{2},........,{\bf r}_{N}/{\bf r}_{1})d{\bf
  r}_{2}.......d{\bf r}_{N}=1 \forall {\bf
  r}_{1}\nonumber\\
{\it (ii)}~~~~~~f({\bf
  r}_{1},..{\bf r}_{i}...{\bf r}_{j-1},{\bf r}_{j+1}...,{\bf r}_{N}/{\bf r}_{j})=0; for~~ i=j;\forall i,j=1,N
\nonumber\\
{\it (iii)}~~~~~~f({\bf
  r}_{1},.....{\bf r}_{i}..,{\bf r}_{j}...{\bf r}_{k-1},{\bf r}_{k+1}..,{\bf r}_{N}/{\bf r}_{k})=0; for~~ i=j;\forall i,j\neq k
\label{condeq1}
\end{eqnarray}
The property ${\it (iii)}$ of Eq.\ref{condeq1} assures us that $f$ reflects the fermionic character of an electronic wavefunction. In fact it says that if {\bf any} two particles are in the same 'state'
''${\bf r}$'' the probability of that specific global configuration is zero. In principle, together with condition {\it (ii)}, this is a way to mimic the antisymmetric character of the fermionic wavefunction since for fermions $|\psi({\bf r}_{1},...{\bf r}_{i},...{\bf r}_{j},...{\bf r}_{N})|^{2}=0; for~~i=j, \forall i,j$.It must be noticed that condition {\it (iii)} is complementary to {\it (ii)}.
With this formalism the term $\left<\psi\left|T+V_{ee}\right|\psi\right>$ can be written as (see Refs.\cite{sears,lui1,lui2}):
\begin{eqnarray}
\left<\psi\left|T+V_{ee}\right|\psi\right>= \frac{1}{8}\int\frac{|\nabla\rho({\bf r})|^{2}}{\rho({\bf r})}d{\bf r}+\frac{1}{8}\int\rho({\bf r})\left[\int_{\Omega_{N-1}}\frac{|\nabla_{{\bf r}}f({\bf r}^{'},....,{\bf r}_{N}/{\bf r})|^{2}}{f({\bf r}^{'},.....,{\bf r}_{N}/{\bf r})}d{\bf r}^{'}....d{\bf
  r}_{N}\right]d{\bf r}+\nonumber\\ \left(N-1 \right)\int\rho({\bf r})\left[\int_{\Omega_{N-1}}\frac{f({\bf r}^{'},.....,{\bf r}_{N}/{\bf r})}{|{\bf r}-{\bf r}^{'}|}d{\bf r}^{'}....d{\bf
  r}_{N}\right]d{\bf r}
\label{eq3}
\end{eqnarray}
where we have identified ${\bf r}_{1}$ with ${\bf r}$ and made use of the property of electron indistinguishability, thus ${\bf r}$ could be identified with any of the ${\bf r}_{i}$ (and the same for ${\bf r}^{'}$ identified here with ${\bf r}_{2}$) without changing the results; a further consequence is that the Coulomb expression (last term on the r.h.s.) is written as the sum of $N-1$ identical terms for the generic ${\bf r}$ and ${\bf r}^{'}$ particles. Using Eq.\ref{eq3} the Levy-Lieb constrained-search formulation can then be written as:
\begin{equation}
E_{0}=Min_{\rho}\left(Min_{f}\left(\Gamma[f,\rho]\right)+\frac{1}{8}\int\frac{|\nabla\rho({\bf r})|^{2}}{\rho({\bf r})}d{\bf r}+\int v({\bf r})\rho({\bf r})d{\bf r}\right)
\label{eq4b}
\end{equation}
where
\begin{eqnarray}
\Gamma[f,\rho]=\frac{1}{8}\int\rho({\bf r})\left[\int_{\Omega_{N-1}}\frac{|\nabla_{{\bf r}}f({\bf r}^{'},....,{\bf r}_{N}/{\bf r})|^{2}}{f({\bf r}^{'},.....,{\bf r}_{N}/{\bf r})}d{\bf r}^{'}....d{\bf
  r}_{N}\right]d{\bf r}+\nonumber\\ (N-1)\int\rho({\bf r})\left[\int_{\Omega_{N-1}}\frac{f({\bf r}^{'},.....,{\bf r}_{N}/{\bf r})}{|{\bf r}-{\bf r}^{'}|}d{\bf r}^{'}....d{\bf
  r}_{N}\right]d{\bf r}.
\label{eq5}
\end{eqnarray}
In this way we have transferred the problem from  from $\psi$ to $f$ which means that the focus is now on $\Gamma[f,\rho]$, i.e.,  as discussed in Ref.\cite{lui2}, the correlation functional.
\section{A practical Example: The parametric exponential form of $f$}
In our previous work \cite{lui2}, we have proposed an approximation for $f$ based on a two-particle factorization:
\begin{equation}
f=\Pi_{i=2}^{N}h_{i}(E_{H}({\bf r},{\bf
  r}_{i}))=\Pi_{i=2}^{N}e^{(N-1){\overline E}({\bf r})}e^{-E_{H}({\bf r},{\bf r}_{i})}
\label{exp1}
\end{equation}
where
\begin{equation}
e^{-{\overline E}({\bf r})}=\int_{\omega}e^{-E_{H}({\bf r},{\bf
    r}_{i})}d{\bf  r}_{i}.
\label{norm}
\end{equation}
here $E_{H}({\bf r},{\bf r}_{i})=\frac{\rho({\bf r})\rho({\bf r}_{i})}{|{\bf r}-{\bf r}_{i}|}$, $N$ is the number of particle, and $\omega$ the volume corresponding to one particle. Such an approximation, due to its simplicity, allows us to write an analytic expression of the Fisher functional which can be used in a straightforward way in numerical calculations. However it does not satisfy the condition ${\it (iii)}$ of Eq.\ref{condeq1}, and, for this reason, in order to use it into the Levy-Lieb constrained-search scheme it must be extended. The expression we propose here is the following:
\begin{equation}
f({\bf r}_{2},...{\bf r}_{N}/{\bf r})=\Pi_{n=2,N}e^{{\overline{\overline E}}({\bf r})-\gamma E_{H}({\bf r},{\bf r}_{n})}\times \Pi_{i>j\neq 1}e^{-\beta E_{H}({\bf r}_{i},{\bf r}_{j})}
\label{examp1}
\end{equation}
with:
\begin{equation}
e^{-{\overline{\overline E}}({\bf r})}=\int\Pi_{n=2,N}\Pi_{i>j\neq 1} e^{-\gamma E_{H}({\bf r},{\bf r}_{n})-\beta E_{H}({\bf r}_{i},{\bf r}_{j})} d{\bf r}_{2}.....d{\bf r}_{N}
\label{examp1b}
\end{equation}
Here $\gamma$ and $\beta$ are two free parameters. As it can be easily verified this expression of $f$ satisfies all the requirements of Eq.\ref{condeq1}. The meaning of $f$ as expressed in Eq.\ref{examp1} is that the probability of finding a certain configuration for the $N-1$ particles, having fixed particle ${\bf r}_{1}={\bf r}$, depends not only on the fixed particle and its interaction with the $N-1$ other particles as before, but also on the mutual arrangements of the $N-1$ particles (it has also to be kept in mind that using the particle indistinguishability the formalism can be applied to any ${\bf r}_{i}$ as a fixed particle). The parameters $\gamma$ and $\beta$ express how important the $N-1$ mutual interactions are with respect to the interactions with ${\bf r}$. Being now $f$ a biparametric function, one can use the Levy-Lieb constrained search in our formulation and find the optimal values for $\gamma$ and $\beta$. This practical example shows two different aspects of our formulation; basically we have shown that indeed it is possible to build a function $f$ and actually it can be chosen in a way that its optimal expression can be determined via the constrained-search formulation.  It must be noticed that this form of $f$ is still rather simple since the spins are not explicitly considered when constructing the function and thus one cannot distinguish between the exchange and the correlation part of the electron-electron interaction as it is done in standard Density Functional Theory; as a consequence one should expect only an overall average description of these two terms which are here incorporated into the global correlation. However the construction of a more complete expression of $f$, which takes care of the effects of the spins, is the subject of current investigation. This emphasizes once more the merit of the general procedure shown here, that is different expressions of $f$, with different degrees of complexity, can be proposed and their relative validity checked by the constrained-search procedure.

\section{Discussion and Conclusions}
As anticipated in the introduction, the consequences of Eqs.\ref{eq4b},\ref{eq5} are rather interesting. The Levy-Lieb variational principle can be reformulated as: {\it The universal functional $F[\rho]$ is the one with the minimum correlation functional with respect to the electron conditional probability density}. 
This new interpretation of the HK universal functional tells us that only an accurate description of the correlation effects, considering the Weizsacker term as a necessary term, leads to an accurate description of the whole energy functional; such a criterion is necessary and sufficient. It is obvious that it is necessary; without knowing $\Gamma[f,\rho]$,
$F[\rho]$ cannot be known; it is sufficient because once $\Gamma[f,\rho]$ or better $f({\bf r}_{2},...{\bf r}_{N}/{\bf r}_{1})$ is (in principle) known than the whole energy functional is known explicitely. Clearly, the ''true'' $f({\bf r}_{2},...{\bf r}_{N}/{\bf r}_{1})$ is very difficult if not impossible to obtain
\cite{kohout}, however it can be sufficiently well described on the basis of mathematical requirements and physical intuition as done for example in Ref.\cite{lui2} and as shown in the previous section. From this point of view, Eqs.\ref{eq4b},\ref{eq5}, can be seen as an objective criterion  to design, on the basis of physical intuition and fundamental mathematical requirements, valid energy functionals. In fact, as done in Ref.\cite{lui2} and in the previous section, one can construct well-founded expressions for $f$ keeping in mind the physical meaning
of the electron correlation effects and the necessary related mathematical prescriptions of Eq.\ref{condeq1}. Next one can make use of Eqs.\ref{eq4b},\ref{eq5} and choose among different functional forms of $f$, the one giving the ''smaller'' $\Gamma$. 
It must be noticed that in this work we do not claim that finding a functional form of $f$ is easier or more rigorous than to find an exchange-correlation functional in standard Density Functional Theory; it represents an alternative or complementary approach to the latter. However, the approach based on $f$ allows one to express in a more direct way, via the choice of different forms of $f$, the  physical principles related to the electron correlation effects and to have an explicit form of the correlation term for the kinetic functional which is of great advantage for kinetic functional based methods (see e.g. Refs.\cite{ofdft1,ofdft2}).An important aspect linked to the statement above is that the 
term: $\frac{1}{8}\int\rho({\bf r})\int_{\Omega_{N-1}}\frac{|\nabla_{{\bf r}}f({\bf r}^{'},....,{\bf r}_{N}/{\bf r})|^{2}}{f({\bf r}^{'},.....,{\bf r}_{N}/{\bf r})}d{\bf r}^{'}....d{\bf
  r}_{N}d{\bf r}$, is the well known {\bf non local Fisher information functional} about which a vast literature is available (see e.g. \cite{sears,nagy,romera} and references therein); this term is very often  linked to the electron correlation functional and electronic properties(see Refs.\cite{new1,new2}), our work further clarifies this connection, suggesting that the results known from the analysis of the Fisher functional could be employed in this context. In conclusion we have shown an alternative view of the Levy-Lieb constrained search approach and provided an example which clarifies the practical advantage of our idea; in this sense the present work it is not merely a
marginal new formal contribution to a rather well-known method, but gives a new powerful insight into the field of applicability for realistic systems.\\

{\bf Acknowledgments}\\
I would like to thank Luca Ghiringhelli for a critical reading the manuscript.


\begin{thebibliography}{9}
\bibitem{au}
We use atomic units where $\hbar$, $e$ and $m$ are equal to one.
\end{thebibliography}
\end{document}